\documentclass{emulateapj}
\usepackage{apjfonts}
\usepackage{epsf}
\usepackage{psfig}
\begin{document}

\slugcomment{Submitted to ApJ}
\shortauthors{J. M. Miller et al.}
\shorttitle{MAXI J1305$-$704}

\title{Chandra Spectroscopy of MAXI J1305$-$704: Detection of an Infalling Black Hole Disk Wind?}

\author{J.~M.~Miller\altaffilmark{1},
  J.~Raymond\altaffilmark{2},
  T.~R.~Kallman\altaffilmark{3}
  D.~Maitra\altaffilmark{1},
  A.~C.~Fabian\altaffilmark{4},
  D.~ Proga\altaffilmark{5}
  C.~S.~Reynolds\altaffilmark{6},
  M.~T.~Reynolds\altaffilmark{1},
  N.~Degenaar\altaffilmark{1,7},
  A.~L.~King\altaffilmark{1},  
  E.~M.~Cackett\altaffilmark{8},
  J.~A.~Kennea\altaffilmark{9},
  A.~Beardmore\altaffilmark{10}
}

\altaffiltext{1}{Department of Astronomy, The University of Michigan, 500
Church Street, Ann Arbor, MI 48109, jonmm@umich.edu}
\altaffiltext{2}{Harvard-Smithonian Center for Astrophysics, 60 Garden
  Street, Cambridge, MA, 02138}
\altaffiltext{3}{NASA Goddard Space Flight Center, Code 662,
  Greedbelt, MD, 20771, USA}
\altaffiltext{4}{Institute of Astronomy, University of Cambridge,
Madingley Road, Cambridge CB3 OHA, UK}
\altaffiltext{5}{Department of Physics, University of Nevada, Las
  Vegas, Las Vegas NV, 89154}
\altaffiltext{6}{Department of Astronomy, University of Maryland,
  College Park, MD, 20742-2421}
\altaffiltext{7}{Hubble Fellow}
\altaffiltext{8}{Department of Physics and Astronomy, Wayne State
  University, Detroit, MI, 48201}
\altaffiltext{9}{Department of Astronomy \& Astrophysics, The
  Pennsylvania State University, 525 Davey Lab, University Park, PA,
  16802}
\altaffiltext{10}{Department of Physics and Astronomy, University of
  Leicester, Leicester, LE1 7RH, UK}


\begin{abstract}
We report on a {\it Chandra}/HETG X-ray spectrum of the black hole
candidate MAXI J1305$-$704.  A rich absorption complex is detected in
the Fe L band, including density--sensitive lines from Fe XX, XXI, and
XXII.  Spectral analysis over three bands with photoionization models
generally requires a gas density of $n \geq 10^{17}~ {\rm cm}^{-3}$.
Assuming a luminosity of $L = 10^{37}~ {\rm erg}~ {\rm s}^{-1}$, fits
to the 10--14~\AA~ band constrain the absorbing gas to lie within $r =
3.9\pm 0.7 \times 10^{3}$~km from the central engine, or about $r =
520\pm 90~ (M/5~M_{\odot})~ r_{g}$, where $r_{g} = GM/c^{2}$.  At this
distance from the compact object, gas in Keplerian orbits should have
a gravitational red-shift of $z = v/c \simeq 3\pm 1 \times 10^{-3}~
(M/5~M_{\odot})$, and any tenuous inflowing gas should have a
free--fall velocity of $v/c \simeq 6\pm 1 \times 10^{-2}~
(M/5~M_{\odot})^{1/2}$.  The best-fit single-zone photoionization
models measure a red-shift of $v/c = 2.6-3.2\times 10^{-3}$.  Models
with two zones provide significantly improved fits; the additional
zone is measured to have a red-shift of $v/c =$4.6--4.9$ \times
10^{-2}$ (models including two zones suggest slightly different radii
and may point to lower densities).  Thus, the shifts are broadly consistent
with the photoionization radius.  The results may be explained in
terms of a ``failed wind'' like those predicted in some numerical
simulations.  We discuss our results in the context of accretion flows
across the mass scale, and the potential role of failed winds in black
hole state transitions.
\end{abstract}

\section{Introduction}
Ionized X-ray absorption lines are now observed in high resolution
X-ray spectra from a fairly large number of X-ray binaries.  In some
cases, the lines are observed close to their rest wavelengths, and
likely arise in the coronal interstellar medium (e.g. Juett et
al.\ 2006) or in ionized gas within the binary (Diaz-Trigo et
al.\ 2006).  When such lines are significantly blue-shifted, however,
they likely trace outflowing disk winds.  Ionized X-ray winds have now
been observed in both stellar-mass black hole systems (e.g. Miller et
al.\ 2006a,b; Kubota et al.\ 2007; Miller et al.\ 2008; Ueda, Yamaoka,
\& Remillard 2009; Neilsen \& Lee 2009; King et al.\ 2012) and in
neutron star binaries (Ueda et al.\ 2004, Miller et al.\ 2011).  These
winds may eject more matter than is able to accrete onto the compact
object, making their study essential to a complete picture of disk
accretion.

Understanding the physical mechanisms by which the winds are launched,
driven, and modulated may be key to revealing the physics of disk
accretion itself (Miller et al.\ 2006a, 2008), as well as black hole
state transitions (e.g. Miller et al.\ 2008, Neilsen \& Lee 2009,
Ponti et al.\ 2012, Miller et al. 2012).  Global studies of disk winds
and their relationship to jet outflows offer one important angle on
this problem (e.g. King et al.\ 2013).  Another angle is available
when X-ray spectra contain density-sensitive lines that can help to
accurately reveal the launching radius and mass outflow rate of the
wind.  In such cases, it is possible to distinguish thermal driving
(e.g. Begelman, McKee, \& Shields 1983) from magnetic driving
mechanisms (e.g. Blandford \& Payne 1982, Proga 2003; also see
Reynolds 2012).

Although He-like triplets and emission line ratios can give gas
densities, emission can originate in regions unrelated to a wind.
This is particularly problematic in the case of AGN, where
illumination of the torus can give rise to He-like triplets that do
not necessarily reflect the state of gas closer to the black hole.
Spectra observed from X-ray binaries, in contrast, typically span a
far narrower range in ionzation, and it is particularly unlikely that
multiple regions may contribute to a line spectrum in systems with low
mass (e.g. solar mass) companions.  Robust wind constraints can be
obtained using the density-sensitive Fe XXII line pair (11.77\AA,
11.92\AA; see Mauche, Liedahl, \& Fournier 2003); blue-shifted
detections in the spectrum of GRO J1655$-$40 permitted strong wind
constraints in that system, and a weak detection of the line pair was
claimed in NGC 4051 (Miller et al.\ 2008; King, Miller, Raymond 2012).
At higher densities, it is possible to use the ratio of Fe XXI lines
(12.259~\AA~ and 12.325~\AA) as a density diagnostics (see, e.g.,
King, Miller, \& Raymond 2012).  Especially in X-ray binaries, the use
of such lines is partly dependent on observing sources with a modest
column density through the ISM, in order to preserve flux in the Fe L
band.

MAXI J1305$-$704 was discovered on 09 April 2012, at a flux of
30~mCrab in the 4--10~keV band (Sato et al.\ 2012).  The low column
density along this line of sight enables optical observations in many
bands, and a counterpart was soon identified (Greiner, Rau, \& Schady
2012).  It is particularly notable that the source colors, optical
spectrum, and hard X-ray rise are very similar to the properties
observed from the well-known black hole XTE J1118$+$480 (Greiner, Rau,
Schady 2012; Charles et al.\ 2012).  MAXI J1305$-$704 was also quickly
detected in radio bands (Coriat et al.\ 2012).  This is also weakly
suggestive of a black hole system, as black holes appear to be more
luminous in radio than neutrons stars (Migliari \& Fender 2006).  Suwa
et al.\ (2012) noted that outburst evolution and states observed from
MAXI J1305$-$704 are also suggestive of a black hole, but cannot
rigidly exclude a neutron star.  The source has been observed to
display X-ray flux dips (see, e.g. Kuulkers, Motta, \& Belloni 2012),
indicating a relatively high inclination, and permitting a binary
period measurement of $9.74\pm 0.04$~hours (Shidatsu et al.\ 2013).
Thus, although there are indications of a black hole, the distance to
MAXI J1305$-$704, and the nature of the compact object are currently
unknown.

A snapshot exposure with {\it Swift} detected an Fe K absorption line
in the persistent emission of MAXI J1305$-$704, as well as some
preliminary indications for absorption in the Fe L band (Miller et
al.\ 2012b).  We then requested a {\it Chandra} observation
with the particular aim of detecting a disk wind and Fe XXII
absorption lines.  A preliminary report on the {\it Chandra} spectrum
is given in ATEL 4191 (Miller et al.\ 2012c).

\section{Observations and Data Reduction}
MAXI J1305$-$704 was observed with {\it Chandra} for 30.0~ks starting
on 2012 August 02 at 12:22:17 (UT).  The incident flux was dispersed
onto the ACIS-S array using the HETG.  The ACIS array was operated in
the ``faint'' data mode as the source was not as bright as many X-ray
binaries, which often require the array to be operated in ``continuous
clocking'' mode.  The observation was executed using the MAXI
coordinates (13h 05m 39.9s, $-$70$^{\circ}$ 24' 54''; Sato et
al.\ 2012).  However, an improved source position was later obtained
in a {\it Swift} imaging observation (13h 06m 55.53s, $-$70$^{\circ}$
27' 01.8''; Kennea et al.\ 2012).  Thus, the true source position is
approximately 6.7' from the position at which {\it Chandra} pointed.
The roll angle of the observatory was fortuitous, however, and the
source was not only imaged on the ACIS-S array, but on the S3 chip.

All of the {\it Chandra}-specific data reduction tasks in this paper
were performed using CIAO version 4.4.1 and the latest associated
calibration database files.  Other tasks were accomplished using the
tools and calibration files available in the HEASOFT suite, version
6.12.  Spectral fits were made using XSPEC version 12.7.1 (Arnaud \&
Dorman 2000).  The spectrum is extremely sensitive so we employed
$\chi^{2}$ fitting statistics and calculated significance levels using
the F-test.  All errors quoted in this work correspond to $1\sigma$
confidence limits.

The zeroth order image of MAXI J1305$-$704 is extended into an
elliptical shape owing to its off-axis position.  We therefore
investigated whether or not the automatic processing used to create
level-2 event files had correctly identified the centroid of the
zeroth order image.  Owing to the fact that this position is used to
set the wavelength grid for the MEG and HEG spectra, the wavelengths
of strong lines in opposing spectral orders (e.g. $+$1 and $-1$) can
be used to check the zeroth order position.  Discrepencies in the
wavelengths of strong lines were found, and so we iteratively adjusted
the position of the zeroth order, produced new event files, spectral
files (using ``acis\_process\_events'', ``tgdetect'',
``tg\_create\_mask'', ``tg\_resolve\_events'', and ``tgextract''), and
responses (using ``mkgrmf'' and ``fullgarf''), until agreement was
found between opposing first-order spectra in both the MEG and HEG.
The ``add\_grating\_orders'' tools was used to combine the first-order
MEG and HEG spectra and responses prior to analysis.  One arm of the
HEG is truncated owing to the offset pointing, and the remaining arm
lacks the sensitivity for line studies in the Fe K band.  The
collecting area of the MEG is higher than the HEG in the Fe L band
even in a nominal observation.  For these reasons, the analysis below
is restricted to the combined first-order MEG spectrum of MAXI
J1305$-$704.

Zeroth order spectra and response files were extracted using the
``specextract'' script.  The entirety of the zeroth order image was
extracted using a large elliptical region.  As expected based on the
source image, fits to this extraction reveal the effects of photon
pile-up, most promininently in the high energy band where
artificially--summed events are registered as single photons.  While
the continuum is affected, the centroid energy of absorption lines is
robust against pile-up, and this later serves as a check on the
results obtained from the dispersed spectrum.

\section{Analysis and Results}
\subsection{Light Curve}
As noted above, MAXI J1305$-$704 is known to be a ``dipping'' source,
based on numerous {\it Swift} observations.  Flux dips are most often
observed in systems in which the line of sight is close to the plane
of the accretion disk, likely signaling that the dips are caused by
material just above the outer disk (e.g. Diaz-Trigo et al.\ 2006).
This line of sight is fortuitous for efforts to study equatorial
winds (Miller et al.\ 2006b, 2008; Ponti et al.\ 2012).  We therefore
extracted and examined the lightcurve of the dispersed spectra.  Using
``dmextract'', we extracted the first-order MEG and HEG events in the
0.3--10.0~keV range, and summed the light curve into 10~s bins for
visual clarity (see Figure 1).  The flux is found to decrease by a
factor of $\simeq$2 in the final few kiloseconds of the observation,
with marginal evidence of a short, weak dip event about 28~ks into the
observation, lasting less than 1~ks.  If the slow decline at the end
of the observation is the start of a dip event, it is much weaker than
the strong, sharp dips reported by e.g. Kennea et al.\ (2012), and
unlikely to have a strong impact on the spectrum.  We therefore
proceeded to analyze the time-averaged spectrum gathered from the
entire {\it Chandra} observation.

\subsection{The First-Order MEG Spectrum}
We considered the combined first-order MEG spectrum of MAXI
J1305$-$704 in the 8--19.5~\AA~ band, and in selected slices of this
range.  This full band is clearly dominated by strong -- and sometimes
blended -- Fe L absorption lines.  At longer wavelengths, the spectrum
is attenuated by interstellar absorption.  At shorter wavelengths,
there are sharp changes in the instrument response owing to the
off-axis nature of the source position, and resulting consequences for
chip gaps and other features.  Though this is a limited band and there
is evidence of absorption in e.g. the 6--8~\AA~ band as well, the
8--19.5~\AA~ band effectively captures the Fe L absorption seen with
{\it Swift}.

The spectral continuum is not the focus of our investigation; we
regard the continuum as largely fiducial as the large offset pointing
may cause flux calibration uncertainties.  However, a few notes are
required.  All spectral fits to the MEG spectrum were made using the
``tbabs'' absorption model with corresponding atomic cross-sections
and abundances (Wilms, Allen, \& McCray 2000).  The effective neutral
hydrogen column density along the line of sight to MAXI J1305$-$704 is
very low and difficult to constrain; however, a value of $N_{H} =
1.5\times 10^{21}~ {\rm cm}^{-2}$ permits a reasonable
characterization of the low-energy spectrum.  This value was fixed in
all fits.  The MEG spectrum is dominated by soft, thermal emission
that can be fit with a $kT \simeq 1$~keV disk blackbody model (Mitsuda
et al.\ 1984).  A hard spectral component is required and was
characterized with a simple power-law model; the spectral index of
this model depends on the band under consideration.  Over the
8--19.5~\AA~ range, a soft index of $\Gamma \simeq 3$ yields good
fits.  This continuum is broadly similar to that observed in GRO
J1655$-$40 when a dense, ionized, magnetically-driven wind was
detected (Miller et al.\ 2006b, 2008; Kallman et al.\ 2009; also see
Neilsen \& Homan 2009 and Neilsen 2013), and also similar to the soft,
disk--dominated continua found as winds were launched in H 1743$-$322
(Miller et al.\ 2006), GRS 1915$+$105 (Ueda et al.\ 2009), and IGR
J17091$-$3624 (King et al.\ 2012).  Last, we found that two broad
($\sigma \simeq 0.1$~keV) Gaussian emission functions with centroids
between 0.7--0.8~keV and 1.2--1.3~keV generally improved the fits.
These lines wer also detected in a {\it Suzaku} spectrum of MAXI
J1305$-$704 (Shidatsu et al.\ 2013).  They are broadly similar to
lines reported in compact X-ray binaries by Madej et al.\ (2010) and
Madej \& Jonker (2011), and interpreted as relativistic disk lines.
We suggest that relativistic O VIII and Fe L lines may provide a
reasonable interpretation of the lines we have detected (also see
Shidatsu et al.\ 2013).  They are adequately fit with Gaussians and we
defer more detailed analysis in order to focus on the rich absorption
spectrum.

Fits to the 8--19.5~\AA~ band with a $kT = 1.0$~keV disk plus
blackbody and power-law model do not give formally acceptable fits,
but do allow for a basic flux characterization.  Using this model, a
flux of $F = 3.1\times 10^{-9}~ {\rm erg}~ {\rm cm}^{-2}~ {\rm
  s}^{-1}$ is implied for the 0.3--10.0~keV band.  The source is not
positioned close to the Galactic Center, but the distance to Galactic
Center is a useful benchmark.  This flux gives a luminosity of $L =
2.7\times 10^{37} (d/8.5~{\rm kpc})^{2}~ {\rm erg}~ {\rm s}^{-1}$.
Thus, even for fairly large distances, the luminosity of MAXI
J1305$-$704 is unlikely to greatly exceed $L = 10^{37}~ {\rm erg}~
{\rm s}^{-1}$, and this value gives a useful standard for eventual
scaling when the distance is known.

Unlike H~1743$-$322 and IGR J17091$-$3624, where only two lines are
detected in the Fe K band, and unlike GRO J1655$-$40 and GRS
1915$+$105, where a high ionization parameter spurred absorption
across the full soft X-ray band, the absorption in MAXI J1305$-$704 is
strongly concentrated in the Fe L band, near to 1 keV (1~keV $\times$
1~\AA~ $=$ 12.39852), though Fe K absorption is also evident.  Thus,
whereas phenomenological modeling of the absorption lines with simple
Gaussian functions was sensible in other sources, the presence of
multiplets and line blends limits the utility of simple Gaussian
functions in describing the spectrum of MAXI J1305$-$704.  Moreover,
in this situation, the transition that one might assign to a given
line can depend on the gas density and ionization, which requires
careful modeling of numerous lines using a self-consistent model.

The density-sensitive Fe XXII lines at 11.77~\AA~ and 11.92~\AA~ are
important exceptions.  The ratio of these lines is a density
diagnostic for $n \geq 10^{12}~ {\rm cm}^{-3}$ (e.g. Mauche, Liedahl,
\& Fournier 2003).  Even visual inspection of the lines confirms that
the 11.92~\AA~ line is stronger than the 11.77~\AA~ line in the
spectrum of MAXI J1305$-$704, in contrast to the ratio found in GRO
J1655$-$40 and NGC 4051 (King, Miller, \& Raymond 2012).  We
constructed a crude power-law model of the continuum in the
11.2--12.7~\AA~ band and characterized the absorption with Gaussian
functions in order to make fits to the Fe XXII lines.  Via the
Gaussian fits, the $11.92~\AA / 11.77~\AA$ flux ratio exceeds 1.6;
CHIANTI simulations presented in Figure 4 of Miller et al.\ (2008)
show that this requires a density of $n > 10^{15} {\rm cm}^{-2}$.

\subsection{Photoionization Modeling}
Owing to the complexity of the observed spectrum, we calculated a
large grid of photoionization models using XSTAR version 221bn15
(Kallman \& Bautista 2001).  Based on our broad-band spectral fits, we
calculated models assuming a simple, unabsorbed, $kT = 1.0$~keV
spectrum incident upon the gas, with a luminosity of $L = 1.0\times
10^{37}~ {\rm erg}~ {\rm s}^{-1}$.  The detection of flux dips in {\it
  Swift} monitoring of MAXI J1305$-$704 demands a high inclination; an
equatorial wind or flow is also indicated by the lack of strong,
narrow emission lines.  Their absence indicates the lack of dense
material above our line of sight that could reprocess the X-rays and
contribute emission lines.  For simplicity and consistency with other
sources (see, e.g., King et al.\ 2013), our XSTAR models assumed a
covering factor of $\Omega/4\pi = 0.5$.  Solar abundances were assumed
for all elements and all models.  Numerous tests indicated that a
turbulent velocity of $v = 700$~km/s gives the best description of the
data, and this was then assumed in all models.

As noted above, when the density-sensitive Fe XXII lines are modeled
with Gaussian functions, their ratio of $\simeq 1.6$ implies a density
of at least $n \geq 10^{15}~ {\rm cm}^{-3}$ (Miller et al.\ 2008).
Nevertheless, in order to test the sensitivity of the spectrum to
different densities, we constructed models with $n = 10^{13},
10^{14}, 10^{15}, 10^{16}, 10^{17}, {\rm and}~ 10^{18}~ {\rm cm}^{-3}$.
The models with $n \leq 10^{15}~ {\rm cm}^{-3}$ produced poor fits to
the spectrum of MAXI J1305$-$704.  In particular, those models
over-predicted many Fe L lines, and predicted Fe XXII lines where the
11.77\AA~ line was much stronger than the 11.92\AA~ line.  We
therefore constructed a grid of 8000 models using a computing cluster,
spanning $2.0 \leq {\rm log}(\xi) \leq 4.0$, $5\times 10^{21}~ {\rm cm}^{-2}
\leq N_{H} \leq 5\times 10^{23}~ {\rm cm}^{-2}$, and $10^{16}~ {\rm
  cm}^{-3} \leq n \leq 10^{18}~ {\rm cm}^{-3}$.

We explored fits over three different wavelength regions, selected to
focus narrowly on the Fe XXI and Fe XXII lines (11.2--13.2\AA), to
include the Fe XXII lines and nearby Fe L shell lines (10.0--14.0\AA),
and the full MEG first-order band considered in this work
(8.0--19.5\AA).  In the narrow bands, the continuum was fit with
phenomenological power-law models; in the last (widest) band, more
physical disk blackbody and power-law models were combined to describe
the continuum.  In each case, the corresponding emission line spectrum
generated by XSTAR was added to the continuum, with $n, N, \xi, {\rm
  and}~ v$ all fixed to the values deterined by the absorption
component.  The addition of the corresponding emission line spectra
does not substantially alter the fits; it is simply a measure taken to
ensure internal consistency.  The results of these fits are given in
Table 1.  The lines that contribute most to the absorption spectrum
are given in Table 2.  Fits to specific wavelength ranges, and
comparisons between specific models, are shown in Figures 2, 3, and 4.

In all fits with a single-zone photoionized absorber, the spectrum
requires a very high density.  In fits to the 10--14~\AA~ region, for
instance, Model 5 ($n = 6\pm 1 \times 10^{17}~ {\rm cm}^{-3}$) is a
$5.7\sigma$ improvement over Model 7 (density fixed at $n = 1.0\times
10^{16}~ {\rm cm}^{-3}$).  Some lines -- especially the cluster near
to 13\AA~ -- appear to require an extremely specific set of
parameters, and are therefore better modeled in fits to a narrow
wavelength range.  The O VIII Ly-$\alpha$ and Ly-$\beta$ lines are
generally fit poorly by the models that fit the Fe L shell lines well.
It is possible that a lower density and higher ionziation absorber can
fit these lines better.  However, fits with a single zone where
log(n)$= 13.0$, although they do push to log($\xi$) $\geq 4$, are
worse than Model 5 (see Table 1) by a margin of several hundred in
$\chi^{2}$.  Therefore, it appears possible that abundance variations
may account for the poor fits, rather than incorrect descriptions of
the density and ionization.  It is also possible that a model with a
large number of absorption zones, or a continuous range of zones,
might be able to fit both the Fe L lines and O VIII lines without
abundance variations.

The photionization models are driven toward very high density values
partly by the need to fit other density--sensitive lines.  Just as Fe
XXII absorption can reveal the presence of the excited fine structure
level $2p~^{2}P_{3/2}$, providing density measurements in the
$10^{14}$ to $10^{15}$ range, the population of the $2p^{2}~^{1}D$ and
$^{1}S$ levels of Fe XXI and the $2p^{3}~^{2}D$ and $^{2}P$ levels of
Fe XX provide diagnostics for densities above $n \geq 10^{17}~ {\rm
  cm}^{-3}$.  The Fe XXI lines (12.259~\AA~ and 12.325~\AA; see Table
2) are more prominent in the spectrum of MAXI J1305$-$704.  Their
utility of Fe XXI lines was initially demonstrated in King, Miller, \&
Raymond (2012), though in that work the sensitivity to
densities below $n = 10^{14}~ {\rm cm}^{-3}$ comes from the
$2p2~^{3}P_{2}$ level, which has a lower critical density than the
$2p2~^{1}D_{1}$ level that is important for MAXI J1305$-$704.  Whereas
fits with Gaussians are unable to determine the relative contributions
of Fe XXI and Ne X Ly-series lines in his range, the XSTAR
photoionization models can do this self--consistently.  Contributions
from the Fe XX and Fe XXI lines listed in Table 2 are partly
responsible for the statistically significant improvements found in in
Table 1, when comparing single--zone models with $n \geq 10^{17}~ {\rm
  cm}^{-3}$ to those with a fixed value of $n = 10^{16}~ {\rm
  cm}^{-3}$.

In order to be sure that the fitting process did not miss subtle
minima that would have permitted a low value for the gas density, we
explored a broad range of fits.  Grids of 25 pairings of column
density ($N_{H} = 5.0\times 10^{21}, 1\times 10^{22}, 5\times 10^{22},
1\times 10^{23}, 5\times 10^{23}~ {\rm cm}^{-2}$) and the ionization
parameter (${\rm log}\xi = 2.5, 3.0, 3.5, 4.0, 4.5$) were fit to the
10--14~\AA~ range, for four different values of the gas density ($n =
10^{13}, 10^{14}, 10^{15}, 10^{16}~ {\rm cm}^{-3}$).  The results of
this check are displayed in Figure 5.  The size of the plotting symbol
in Figure 5 encodes the change in the fit statistic,
$\Delta(\chi^{2})$, for each of the 100 different points considered.
Clearly, plausible models with lower density simply do not give fits
as good as models with high values of the gas density.

The distance between the central engine and the absorbing gas is
contained in the definition of the ionization parameter, $\xi = L / n
r^{2}$.  If we take the model for the 10--14\AA~ range as
represenative, $r = 3.9\pm 0.7 \times 10^{8}~ {\rm cm}$ or $r = 3.9\pm
0.7 \times 10^{3}~ {\rm km}$.  Nominally, this radius does not rigidly
exclude a geometry like the boundary layer on an accreting white
dwarf; however, this explanation of the absorption region is unlikely
given the similarity of optical spectra of MAXI J1305$-$704 to optical
spectra of confirmed black holes such as XTE J1118$+$480 (e.g. Charles
et al.\ 2012).  This distance corresponds to $r = 1300\pm200~
GM/c^{2}$ assuming that the compact object is an $M = 2.0~M_{\odot}$
neutron star, or $r = 520\pm 90~ GM/c^{2}$, assuming an $M = 5
M_{\odot}$ black hole primary.

In each wavelength range, we also considered fits with two absorbing
zones (see Table 1, and Figures 3 and 4).  These models yielded
enormous statistical improvments.  In fits to the 10--14~\AA~ range,
for instance, the best--fit two--zone model (Model 8 in Table 1) is an
enormous ($>>8\sigma$) improvement over the best-fit single--zone
model (Model 5).  A direct comparision of these models is shown in
Figure 3.  The specific effect of the velocity shifts is shown in
Figure 4.  Most notably, in each wavelength range, the second
absorption zone is more strongly red-shifted, with values approaching
$v/c = 0.05$ (see Table 1).

The XSTAR code we have used does not include the recent results on the
suppression of dielectronic recombination at high densities by Nikolic
et al.\ (2013).  The recombination rates of the Fe L-shell ions are
reduced by factors of $\leq2$ at densities above $10^{17}$ $\rm
cm^{-3}$, and since the ionization parameter determined from the
spectral fits is determined largely by the relative fractions of these
ions, the ionization parameter $\xi$ is even smaller than the values
given by the fits.  That would imply that the values of $r$ quoted in
Table 1 should be increased by a factor of $\sqrt(2)$; however, this
is easily within the uncertainty in the X-ray luminosity.

Several effects can potentially contribute to the breadth of the
absorption lines in the spectrum of MAXI J1305$-$704, including a
degree of instrumental broadening (see below).  Given the small
absorption radius inferred from even single--zone fits, it is
interesting to consider whether or not tangential velocities have been
encoded into absorption.  Doppler--split line pairs are not typically
seen in X-ray absorption as the central engine is taken to be a point
source compared to the absorbing region.  Simple arguments show that
shifts of $v/c \simeq h \sin(\theta) / w^{3/2}$ are expected (where $h$
is the size of the hard x-ray central engine, $w$ is the size of the
wind launching region, and $\theta$ is the inclination angle measured
relative to the disk normal).  Standard disk models predict that
emissivity drops sharply with radius (Shakura \& Sunyaev 1973), and
new work suggests that coronae are also centrally--concentrated
(e.g. Reis \& Miller 2013).  Taking $h = 10 GM/c^{2}$ and $\theta
\simeq 60^{\circ}$, shifts of $v = \pm 400$~km/s are expected.  This
is broadly consistent with the width of the lines detected in MAXI
J1305$-$704, and also consistent with the turbulent velocity assumed
in our XSTAR models ($v = 700$~km/s).  The nominal value of $N/n =
\Delta(r)$ for the single zone models in Table 1 can be as small as 1
km.  It is possible that in the limit of zero intrumental broadening
due to the pointing offset (see below), or at higher spectral
resolution (such as that available with {\it Astro-H}, Takahashi et
al.\ 2012), Doppler-split line pairs would be detected.

However, numerous instrumental and astrophysical effects may serve to
prevent Doppler-split lines, and/or their detection, even in this
spectrum.  Artificial line broadening due to the telescope pointing
offset may have served to blur away the velocity structure.  Optical
depth effects can spoil line splitting signatures.  Very small values
of $\Delta(r)$ partly rely on the assumption of a unity filling
factor; if the gas is sparse and clumpy -- as might be the case if it
is affected by radiation pressure -- values of $\Delta(r)$ could be
much larger.  Moreover, this analysis has only focused on a very small
wavelength region, dominated by transitions to and from the Fe L
shell, and thus centered on a narrow range in ionization.  An
excellent HEG specrum in the Fe K band could have detected absorption
from very highy ionized gas that would have implied a much larger wind
depth.  The two-zone models detailed in Table 1 may already supply a
partial solution: Models 4, 8, and 12 give radii that differ by an
appreciable factor, and strongly suggest that $\Delta(r)/r$ is not
very small.  In this regime, Doppler-split lines from larger radii would
fill-in pairs from smaller radii, and spoil the signature of rotation.

While the continuum shape and flux of the zeroth-order spectrum are
affected by pile-up, the centroid energy of discrete atomic features
within the ACIS S3 CCD spectrum should be relatively unaffected, and
absorption in the Fe K band can serve as a check on the need for a
more red-shifted component in the absorption spectrum.  We therefore
fit the zeroth order ACIS-S3 spectrum of MAXI J1305$-$704.  A clear
broad absorption trough -- likely composed of Fe XXV and Fe XXVI lines
-- is clearly present, consistent with the prior {\it Swift} spectrum
of MAXI J1305$-$704.  In order to focus the action of the XSTAR
photoionization models on the Fe K region, we restricted our fits to
the 3.0--10.0~keV band.  The spectrum suffers from pile-up, and while
a disk blackbody model gives temperatures near to 1~keV, an additional
power-law component with a negative spectral index.  Thus, at least
the high energy part of the spectral model must be considered
fiducial.

The broad absorption feature in the zeroth-order spectrum can be fit
with a single Gaussian function, with a centroid energy of E$=6.73\pm
0.02$~keV and a FWHM of $0.38\pm0.04$~keV.  This nominally indicates a
blue-shift, but this may be artificial.  It is very difficult to
produce only Fe XXV or Fe XXVI in isolation (see, e.g. Kallman \&
Bautista 2001), and this line is too broad to be composed of a single
line, if the feature bears any relation to the lines in the MEG
spectrum.  Fits to these two lines with a single Gaussian function --
even if the two are slightly red-shifted, can result in a net
blue-shift.  The relative strength of Fe XXV and Fe XXVI lines cannot
be determined using Gaussian models for the zeroth-order spectrum, and
many different line flux ratios could be assumed.  It is notable that
a simple assumption -- two Gaussians required to maintain the energy
offset measured between Fe XXV and XXVI in the laboratory, with line
widths required to be equivalent -- yields red-shifted lines at
E$=6.62\pm 0.02$~keV and E$=6.89\pm 0.02$~keV. This is not proof of a
red-shift in the zeroth-order data, but it serves to illustrate a
potential consistency even via phenomenological fits."

The results of several fitting experiments are given in Table 1 and
shown in Figure 6.  As an initial baseline, we fit the spectrum with a
single-zone absorber, with parameters tied to those of the less
shifted component revealed in the MEG spectrum over the 8--19.5~\AA~
band.  Adding a second absorber then gives an improvement significant
at the $6\sigma$ level of confidence based on an F-test (Model 15
versus Model 13 in Table 1), and requires a red shift of $v/c =
0.022(1)$.  We then made fits to the spectrum with a single-zone
absorber, with only the density fixed (since this cannot be
constrained directly in the Fe K band).  Adding a second absorption
zone is then significant at the $4.2\sigma$ level of confidence (Model
16 versus Model 14 in Table 1), and a red-shift of $v/c = 0.021(3)$ is
measured for the second zone.  These are necessarily limited fitting
experiments, and CCD-resolution spectra are not as suited to this
purpose as dispersed spectra.  However, these results do suggest that
a second, more highly red-shifted absorption zone may alse be
indicated in the Fe K band.

\subsection{Ray tracing}
The off-axis telescope does serve to broaden the zeroth-order image,
resulting in reduced resolution in the dispersed X-ray spectrum.  In
addition, it is possible that the offset has induced artificial
red-shifts through distortions to the wavelength grid.  To understand
the effects of the large offset pointing on the observed spectrum, we
ran simulations using version 4.0 of MARX.  This package contains
detailed, CIAO--independent characterizations of the {\it Chandra}
mirrors, detectors, and gratings instruments.  Photons from any input
spectral form can be traced through their interactions with the
telescope and detectors, and compiled into a simulated data file.
Regular CIAO tools can then be run on the simulated data, and
simulated data products and spectra can be compared to the observed
spectra.

The actual pointing position of the {\it Chandra} observation was used
in the MARX simulation.  The true source position was placed off-axis
in the simulation, again exactly as per the actual observation.  Even
the spacecraft roll angle was selected to match the actual
observation.  A simple power-law input spectrum with three strong
Gaussian absorption lines (0.900~keV, 1.000~keV, 1.100~keV, each with
$\sigma = 5.0\times 10^{-3}$~keV) was simulated in XSPEC at very high
resolution, and then normalized to have the same photon flux as the
actual {\it Chandra} observation.  Assuming a point source for MAXI
J1305$-$704, MARX passed rays from this distribution through models of
the HRMA, HETG, and ACIS-S array.  The resultant event list was
reduced in exactly the manner that actual data would be reduced, using
the same CIAO tools.

Fitting the MEG$-$1 and MEG$+$1 orders extracted from simulated data
separately, we find no evidence of a mismatch in the wavelength grid.
Nor was evidence of a net red-shift found.  Indeed, the centroid of
each line in the simulated spectra was found to match the input value
to within 0.001~keV.  (Here it is worth noting that an incorrect
zeroth order position in deriving the real and simulated spectra would
result in line broadening or splitting, not in a net shift.)  However,
there is evidence of line broadening due to the offset pointing, and
the effect may be more pronounced at shorter wavelengths.  Fits to the
simulated MEG$-$1 spectrum, for instance, gave values of $\sigma =
6.2(2)\times 10^{-3}$~keV, $\sigma = 7.4(2)\times 10^{-3}$~keV, and
$\sigma = 8.1(2)\times 10^{-3}$~keV, for the lines inserted at
0.900~keV, 1.000~keV, and 1.100~keV, respectively (recall that
$\sigma_{\rm input} = 5.0\times 10^{-3}$~keV).

In summary, simulations with MARX suggest that the observed lines are
likely to be artificially broadened to some degree, perhaps by
20--50\%, but that the observed red-shifts are likely astrophysical.
This possibility of wavelength--dependent artificial line broadening
is in some conflict with the observed spectrum of MAXI J1305$-$704:
lines at shorter wavelengths do not appear to be systematically
broader than lines at shorter wavelengths.  It is possible that
astrophysical complications, such as a range of ionization parameters,
distances, and turbulent velocities, may have served to obscure any
wavelenth--dependent instrumental broadening.  Although the simulated
spectrum does not show a net red-shift, our view is that an
instrumental origin remains as a remote possibility.  MARX is
extremely powerful, but it is only as good as the intrument
characterizations, and it is possible that our observation pushes
beyond the data on which MARX is based.

\section{Discussion}
We have analyzed a {\it Chandra} high-resolution X-ray spectrum of the
transient X-ray binary MAXI J1305$-$704 in outburst.  A snapshot CCD
spectrum of the source obtained with the {\it Swift}/XRT detected Fe K
absorption, and also suggested strong Fe L shell absorption.  The {\it
  Chandra} spectrum dramatically confirms this suggestion, revealing
an especially complex and rich array of absorption lines from Fe
XIX--XXIV, as well as H-like and He-like charge states of O and Ne,
and likely also Na and Mg (see Tables 1 and 2, and Figures 2, 3, and
4).  The MEG spectrum is extremely sensitive and complex; the rich
wind absorption spectrum detected in GRO J1655$-$40 may be the most
natural point of comparison.  In both cases, the situation is more
similar to fitting UV spectra from quasars rather than typical X-ray
binary spectra: formally acceptable fits with self-consistent
photoionization models are made impossible by the quality of the data
(see, e.g., Miller et al.\ 2006, 2008; Kallman et al.\ 2009; Neilsen
\& Homan 2012).  None of the models presented in Table 1 are formally
acceptable.  However, the relative quality of different models can
still be evaluated via changes in the $\chi^{2}$ fitting statistic.

Most importantly, the density--sensitive Fe XXII line pair at
11.77~\AA~ and 11.92~\AA~ is detected (see Figures 3 and 4), and the
flux ratio of these lines requires a very high density.  Even a crude
comparison to prior calculations suggests $n > 10^{15}~ {\rm cm}^{-2}$
(see, e.g., Miller et al.\ 2008).  For an assumed luminosity of $L =
10^{37}~ {\rm erg}~ {\rm s}^{-1}$ and an ionization parameter of $\xi
\simeq 2$, this density limit requires $r \leq 10^{10}~ {\rm
  cm}^{-2}$.  Detailed fits with a very large and finely-spaced set of
XSTAR photoionization models finds $n \geq 10^{17}~ {\rm cm}^{-2}$ and
$r \simeq 3.9\pm 0.7 \times 10^{8}~ {\rm cm}$ (see Model 5 in Table
1).  Other density-sensitive lines from Fe XX and Fe XXI contribute to
the density determination in fits with photoioniozation models.
Numerous details and possibilities are explored below, but the most
sound and important result of this analysis is that X-ray absorption
has been detected within approximately 5000~km from a compact object
that is likely a black hole.

It is nominally possible that the source harbors a white dwarf and
that the spectrum originates in some sort of photosphere; however,
this is extremely unlikely.  The temperature of the continuum emission
is far more consistent with accretion onto a neutron star or black
hole; so too is the hard X-ray component detected in MAXI J1305$-$704.
No X-ray bursts nor pulsations have been reported in observations of MAXI
J1305$-$704, nor has timing behavior characteristic of many neutron
star X-ray binaries (e.g. strong low--frequency QPOs.).  If the source
is a neutron star, Model 5 constrains the absorption to occur within
$r = 1700~ r_{g}$ for a neutron star with $M = 2~M_{\odot}$.  This is
already very small, even by the standards of X-ray winds in binaries.
However, the optical spectrum and X-ray flux changes observed in MAXI
J1305$-$704 are suggestive of a black hole primary.  In that case, the
absorption originates at or within $520\pm 90 (M / 5~M_{\odot})~ r_g$.
The disk wind in GRO J1655$-$40 is also dense, and likely launched at
similar radii (based on fits to the Fe K band in Miller et al.\ 2008,
as well as models including a higher velocity component in Kallman et
al.\ 2009).  The disk wind in H 1743$-$322 may originate at similar
radii, though the constraints allow for larger radii (Miller et
al.\ 2006).  If the speed of the ``ultra-fast'' outflow found in IGR
J17091$-$3624 reflects its local escape velocity, then it too
originates at approximately $10^{3}~ r_{g}$ (King et al.\ 2012).

At such small launching radii, gravitational red-shifts are
non-negligible.  Appoximating the shift as $v/c = z \simeq GM/rc^{2}$
and taking r $=$ 3000--8000~km (see Models 1, 4, 5, 8, 9, and 12 in
Table 1), $z =$0.9--2.5$\times 10^{-3} M_{5}$ is expected.  Red-shifts
that are broadly consistent with this range are measured from each of
the models in Table 1 with variable density values.  This could be
explained in terms of gas that is executing orbital motion relatively
close to the black hole, that has been lifted above the plane of the
disk, but is not escaping.

When two-zone models are explored, vastly better fits are achieved,
and a much larger red-shift is measured in the additional component.
Models 4, 8, and 12 include zones with $v/c = z = 4.6-4.9\times
10^{-2}$.  It is worth noting that shifts this large could affect the
density constraints obtained from the Fe XXII line pair, because the
11.77~\AA~ line from the more shifted zone can begin to contribute to
the 11.92~\AA~ line from the less shifted zone.  In Models 4 and 8,
good constraints were still achieved for both zones, and $n \geq
10^{17}~ {\rm cm}^{-3}$ is required.  In Model 12, however, a value
consistent with $n = 10^{16}~ {\rm cm}^{3}$ is measured.  Model 8 has
the advantage of applying to a fairly large wavelength range while
still concentrating on the Fe XXII lines, but Model 12 may represent
the best overall description of the spectrum (this is certainly true
in terms of $\chi^{2}$).  This larger shift seen in the second zone in
these models is broadly consistent with the free--fall velocity at the
radius required by photoionization modeling.  Taking $M_{BH} = 5~
M_{\odot}$ and $r = 520~ r_{g}$, for instance, $z = 6.2\times 10^{-2}$
is expected.  The gas need not fall from infinity; traversing a
distance comparable to the absorption radius would give a shift
similar to the full local free--fall velocity.  The fact that the
measured shifts may be slightly below the full free--fall velocity may
support a picture wherein gas has travelled only locally.

Figure 7 shows that these shifts -- if they reflect a gravitational
red-shift and free--fall velocity -- can be realized over a common set
of radii for plausible compact object masses.  This is only marginally
possible for a neutron star with $M_{NS} = 2.0~ M_{\odot}$, but easily
possible for black holes fo a small mass.  The results of fits to the
spectrum of MAXI J1305$-$704 with XSTAR photionization models do not
require that the two components originate at exactly the same radius
in Model 12, but the two zones have overlapping radii for Models 4 and
8 (see Table 1).

Early observations of MAXI J1305$-$704 fueled speculation that the
source may have a short orbital period, and so too have potential
similarities with the short-period black hole XTE J1118$+$480
(e.g. Charles et al.\ 2012).  Shidatsu et al.\ (2013) suggest a
relatively short binary period of $9.74\pm 0.04$~hours, based on an
examination of flux dips.  This would imply a short binary separation,
and would help to explain the small absorption radius and high gas
density implied by the data.  The column density to MAXI J1305$-$704
is quite low, as noted above, and it is reasonable to expect that its
period will be determined via optical methods in the near future.

``Failed winds'' have been predicted in some simulations of accretion
disks and winds in AGN, and they provide a natural explanation for our
results (e.g. Proga \& Kallman 2004, Proga 2005).  Radiation pressure
can launch winds when the Eddington fraction is sufficiently high and
when the gas is self--shielding (this keeps the ionization low and
enhances the force multiplier).  However, if the force multiplier
and/or Eddington fraction are too low, the gas cannot escape.  Indeed,
flows with inward velocity components -- inflows -- have been detected
in the wind region in such simulations (Proga \& Kallman 2004, Proga
2005).

If a ``failed wind'' has been detected in MAXI J1305$-$704, it may not
be exactly like the flows predicted by recent simulations.  Those
studies examined accretion in AGN, wherein the disk emits most of its
energy in the UV.  Although the wind launching region in MAXI
J1305$-$704 would correspond to a region emitting mostly in EUV if
the disk follows the standard $T \propto r^{-3/4}$ prescription, the
force multiplier is low for ${\rm log}(\xi) \simeq$2--3, and this may
mean that magnetic pressure is necessary even to lift the gas into the
line of sight.  As noted by Miller et al.\ (2008) in an analysis of
the wind in GRO J1655$-$40, simulations predict that there should be
ample magnetic flux for this purpose (e.g. Miller \& Stone 2000).  

It is interesting to note that the radius implied by photoionization
modeling is commensurate with the radius at which warps may form.  It
is possible that a warp may help to lift gas up into the line of
sight, or to shield gas from part of the flux from the central engine.
At smaller radii, gravito-magneto-hydrodynamics is expected to anchor
the disk in the plane perpindicular to the black hole spin axis, and
to keep the disk flat (e.g. Bardeen \& Petterson 1975; McKinney,
Tchekhovskoy, \& Blandford 2012).  At larger radii, however,
theoretical treatments suggest that warps can form and may be
sustainable.  Such warps might be expected to give rise to QPOs, and a
connection between ionized absorption and QPO phase was detected in H
1743$-$322 (Miller et al.\ 2006).  The frame time of the ACIS array is
not suited to the detection of even slow QPOs in our observation,
unfortunately, and we defer a timing analysis to later work on
monitoring observations with {\it Swift}.

Failed winds may drive an emerging anti-correlation between winds and
jets, by interfering with the formation or maintenance of a hard X-ray
corona.  Observations clearly signal that disk winds are observed in
soft, disk--dominated states in which the jet is quenched (Miller et
al.\ 2006, 2008, 2012; Neilsen \& Lee 2009; King et al.\ 2012; Ponti
et al.\ 2012).  In constrast, compact, steady jets are ubiquitous in
low--flux, spectrally--hard states wherein coronal emission dominates
disk emission.  As noted by Proga (2005), infusing the region above
the disk with too much cool gas can interfere with a hot,
optically--thin, Comptonizing corona.  Connections between X-ray flux
and radio flux are typically corona--jet connections, not true
disk--jet connections, since X-ray monitors do not typically detect
the cool disk emission that is found in the low/hard state (see, e.g.,
Reis, Miller, \& Fabian 2010).  In the one case where the role of the
disk has been examined directly, it is found not to play a strong
role, and the results suggest that the corona is the base of the
relativistic radio jet (Miller et al.\ 2012d, also see Reis \& Miller
2013).  Winds that flow inward might have a profound effect on the
corona, and thus the base of a would--be jet.

A shortcoming of this scenario stems from the issue of ionization.  It
is possible that magnetic pressure is needed for a failed wind, just
as it is required to drive the wind seen in GRO J1655$-$40 (Miller et
al.\ 2006, 2008; Kallman et al.\ 2009; Neilsen \& Homan 2012; Neilsen
2013).  Given that jets are commonly thought to be powered via
magnetic fields that tap the rotation of the disk (or the spin of the
black hole, or both), given that some winds may be magnetic (as
deduced via a small launching radius, or a mass outflow rate exceeding
the expected range for thermal winds, or both), and given recent work
showing that winds and jets might be powered and regulated in a common
fashion (King et al.\ 2013), it is possible that radiation pressure
does not enter into the disk--wind--jet connection.  A more modest
conclusion is that radiation pressure may be unimportant in X-ray
binaries.  This also follows from the high ionization parameters
observed in X-ray binaries, and simulations that show no force
multiplier above log($\xi$) $\simeq 3$ (e.g. Proga 2003).  In their
study of GRO J1655$-$40, Miller et al.\ (2008) note that magnetic
pressure from MRI processes may be a more likely driver than
magnetocentrifugal forces (also see Neilsen 2013), and that the
wind/jet dichotomy may be driven by changes between a primarily
toroidal field in disk--dominated soft states, and a primarily
poloidal field in corona--dominated hard states with relativistic
jets.

It is possible that other physical scenarios could give the observed
absorption spectrum and velocity shifts.  For instance, the more
strongly shifted, potentially infalling gas could originate in the
outer disk.  If the accretion stream onto the outer disk overflows, as
per some observations and simulations of interacting binaries
(e.g. Armitage \& Livio 1996), some gas may fall inward relatively
independently of the accretion disk.  van Peet et al.\ (2009)
discussed an overflow scenario like this, as a possible explanation
for the Fe L absorption detected in the neutron-star X-ray binary EXO
0748$-$676.  In this case, dips might originate when material still in
the outer disk -- close to the point of impact -- passes across the line
of sight; absorption seen in steady phases (as with this observation
of MAXI J1305$-$704 and some phases of EXO 1748$-$676) would arise via
stream material that had fallen into the inner disk.  If this is
correct, the broadly similar properties of the potentially
free-falling gas in MAXI J1305$-$704, and the gas that might be partly
rotationally--supported, would be largely coincidental.  Moreover,
Armitage \& Livio (1996) note that overflow would be less likely when
the disk is hot and vertically extended, as per a sitation that might
arise when the accretion rate is high and irradiation is important.
The hot disk temperature indicated in MAXI J1305$-$704 indicates a
rather high Eddington fraction, which is always accompanied by
irradiation (Reynolds \& Miller 2013).  For these reasons, we favor a
failed wind scenario.

Although a fairly simple and self--consistent picture can be developed
to explain the data, and though the red--shifts observed are natural
to the radius implied by photoionization modeling, the red-shifts must
be regarded skeptically.  These shifts have not previously been
detected at high statistical confidence in any accretion--powered
system.  Moreover, this observation differs from others in that it has
a large pointing offset, and it is possible that the shifts are
instrumental.

We attempted to test this possibility by simulating this observation
using the {\it Chandra}/MARX ray-tracing packgage.  Those simulations
suggest that the red-shifts are not instrumental in origin; however,
MARX is only as good as the instrument data and characterizations that
drive it.  If its assumptions are incorrect, or if they are limited in
their scope, then MARX may not deliver the correct results for our
observation of MAXI J1305$-$704.

We also attempted to make simple tests of the red-shifts by analyzing
the zeroth-order ACIS-S3 spectrum, which should not be affected in the
same manner as the dispersed spectra.  Again, the test suggests that
the red-shifts are not instrumental (see Table 1), but CCD spectra have
only moderate resolution and the results are not definitive.  Thus, an
instrumental origin for the observed shifts remains a remote
possibility.  Fortunately, the radius implied by photoionization
modeling, and requirement of a high gas density, are unaffected by any
issues related to velocity shifts.  So too, then, is the potential
association of this gas with a failed wind, and the implications of
failed winds for jet production and state transitions.

The neutral ISM along the line of sight might, in different
circumstances, have provided a natural check of the wavelength
calibration of this spectrum.  However, the column density along the
line of sight to MAXI J1305$-$704 is very low (we adopted $N_{H} = 1.5
\times 10^{21}~ {\rm cm}^{-2}$), and this means strong ISM lines will
not be imprinted on the spectrum.  Cases where strong ISM lines have
been observed typically have measured column density values that are a
factor of a few higher (e.g. Pinto et al.\ 2013).  The absorption
spectrum detected within MAXI J1305$-$704 is then approximately an
order of magnitude greater in column density than the ISM along this
line of sight (see Table 1).   The wind column dominated the
potential ISM column and effectively spoiled a natural wavelenth
calibration.  Among the non--Fe lines that are detected and not
clearly blended, including the O VIII Ly-$\alpha$ and Ly-$\beta$
lines, and the Ne IX He-$\alpha$ line, fits with simple Gaussians find
the same $v/c \simeq 3\times 10^{-3}$ shift that is measured in
photoioniztion models dominated by Fe L transitions.

If the red-shifts detected in MAXI J1305$-$704 are astrophysical, they
are exceptional.  It is worth noting, though, that MAXI J1305$-$704 is
exceptional in other ways as well, and that such shifts may be somewhat
less remarkable in this source.  Other X-ray binares, and other
``dippers'', have also shown interesting X-ray absorption spectra.
But as noted by Diaz-Trigo (2006), the gas is typically very highly
ionized, with the most pronounced effects in the Fe K band.  Among
X-ray binaries, MAXI J1305$-$704 is rare in showing such a rich and
complex absorption spectrum in the Fe L band, whether in or out of a
dip phase (also see the spectrum of EXO 0748$-$676 analyzed by van
Peet et al.\ 2009).  Other sources viewed at high inclination show
substantially different spectra, dominated by emission lines; the
differences might be dictated by exactly how high the inclination is.
Her X-1, for instance, shows a complex emission line spectrum
(e.g. Jimenez-Garate et al.\ 2002).  So too do other so-called
``accretion disk coronae'' sources, including e.g. 2S 0921$-$63
(Kallman et al.\ 2003).  Future observations of MAXI J1305$-$704 and
other black hole candidates with {\it Chandra}, {\it XMM-Newton}, and
eventually {\it Astro-H} (Takahashi et al.\ 2012) can help to reveal
the dynamics and physics of wind and jet production and to confirm
possibilities such as red-shifts and Doppler-split line pairs.


\hspace{0.1in}

We gratefully acknowledge comments from the anonymous referee that
improved the clarity and content of this paper.  We thank Harvey
Tananbaum and {\it Chandra} for executing this observation.  We are
indebted to Mateusz Ruszkowski for lending computing cluster clock
cycles to this project.  We acknowledge Mike Nowak, David
Huenemoerder, John Davis, John Houck, Norbert Schulz, and Jonathan
McDowell for helpful discussions.  JMM gratefully acknowledges support
from the {\it Chandra} Guest Observer Program and {\it Swift}.  ND is
supported by NASA through Hubble Postdoctoral Fellowship grant number
HST-HF-51287.01-A from the Space Telescope Science Institute.  APB
acknowledges funding for {\it Swift} at the University of Leicester by
the UK Space Agency.

\clearpage

\begin{table}[t]
\caption{Photoionization Modeling Results}
\begin{tiny}
\begin{center}
\begin{tabular}{lllllllll}
\tableline
\tableline
Model & Range & Notes & $\chi^{2}/\nu$ & $n$ & $N_{H}$ & log($\xi$) & $v/c$ & r \\
  ~   &    ~    &   ~   &     ~          & $(10^{17}~ {\rm cm}^{-3})$ & $(10^{22}~ {\rm cm}^{-2})$ &  ~ & $(10^{-3})$ & $(10^{8}~ {\rm cm})$ \\
\tableline
1  &  11.2--13.2~\AA & 1-zone, free $n$ & 868/384 & 4(2)  & 0.80(6) &  2.03(2) & 3.0(2) & 5(2) \\
2  &    ~           & 1-zone, fixed $n$ & 878/385 & 0.5 &  0.75(5) &  2.05(3) &  2.9(2) & 13(1) \\
3  &    ~           & 1-zone, fixed $n$ & 896/385 & 0.1 & 0.66(5) & 2.05(3) & 2.7(3) & 30(1) \\
4  & 11.2--13.2~ \AA & 2-zone, free $n$ & 715/380 &  2.1(6) & 1.3(4) & 2.60(7) & 2.8(3) & 4(1) \\
  ~ &     ~       &   ~    &    ~    &  5(2)   & 0.67(5)   &  $2.00^{+0.01}$  & 47(1) & 4.7(9)  \\
\tableline
5  &   10.0--14.0~\AA & 1-zone, free $n$ & 1647/785 & 6(1) & 1.05(3) & 2.05(2) & 2.8(2) & 3.9(7)   \\
6  &       ~        & 1-zone, fixed $n$ & 1683/786 & 0.5 & 0.97(3)  & 2.08(1) & 2.3(2) & 13(1) \\ 
7  &      ~        & 1-zone, fixed $n$ & 1716/786 & 0.1 & 0.85(3) & 2.07(2) & 2.1(2) &  29(1) \\ 
8  &   10.0--14.0~\AA & 2-zone, free $n$  & 1434/779 & 2.2(4) & 1.27(5) & 2.60(2) & 2.3(2) & 3.4(7)  \\
  ~ &     ~        &       ~           &    ~     &  7(2)  &  0.70(4)  &  2.00$^{+0.04}$  &  47.0(3) &  3.9(7)  \\
\tableline
9   & 8.0-19.5~ \AA & 1-zone, free $n$ & 4537/2286 & 1.1(1) & 1.22(4) & 2.36(1) & 2.8(2) & 6.3(6) \\
10 &      ~       & 1-zone, fixed $n$ & 4552/2287 & 0.5   &  1.20(3)  &   2.36(1)  &  2.5(2) & 9.3  \\
11 &       ~       & 1-zone, fixed $n$ & 4588/2287 & 0.1   &  1.08(3)  &   2.37(1)  &  2.2(1) & 20.9  \\
12 & 8.0-19.5~ \AA & 2-zone, free $n$ &  3378/2280 & 0.7(2) &   1.10(5)     &   2.62(1)    &  1.8(2) &   6(2)    \\
  ~ &     ~       &        ~         &      ~     &  $0.10^{+0.05}$  &  0.55(5)    &   2.65(1)    &   48(1)  & 15.0   \\
\tableline
13 & 3.0-10.0~keV  & 1-zone, fixed & 592/474 & 0.7 & 1.1 & 2.62 & 1.8 & 5.9 \\
14 &       ~       & 1-zone, free  & 536/471 & 1.0 & 6(2) & 4.0(1) & $<$2 & 1.0 \\
15 & 3.0-10.0~keV   & 2-zone, mixed & 534/471 &  0.7 & 1.1 & 2.62 & 1.8 & 5.9 \\
  ~ &     ~        &        ~      &    ~    &  5  & 1.2(4) & 4.4(1) & 22(1) & 0.3 \\
16 & 3.0-10.0~keV   & 2-zone, free & 509/468 & 1.0 & 6(2) & 4.0(1) & $<$2 & 1.0 \\
  ~ &     ~        &        ~     &    ~    & 1.0 & 6(2) & 4.3(1) & 21(3) & 0.7 \\
\tableline
\tableline
\end{tabular}
\vspace*{\baselineskip}~\\ \end{center} 
\tablecomments{The table above details the results of fits with a
  large grid of XSTAR photoionization models (see the text for
  additional details).  Single-zone fits, models with a fixed
  (artificially low) density, and two-zone models for the combined
  first-order MEG spectrum were considered on three different
  wavelength ranges.  The results of fits to the zeroth-order ACIS-S3
  CCD spectrum on the 3--10~keV band are included in the final segment
  of the table.  The errors listed in the table are $1\sigma$
  confidence intervals.  Symmetric errors are given in parentheses.
  Where errors are not given, the parameter was fixed.  Positive
  velocities indicate a red-shift.  Single-zone fits clearly prefer a
  high density, and the addition of a second zone gives a
  significantly improved fit over all three fitting ranges.  Errors
  are not quoted on radii derived via fits to the ACIS-S3 spectrum
  because the density was fixed in those models.  None of the models
  above provide formally acceptable fits in a statistical sense,
  similar to the situation in complex optical and UV spectra, and in
  fits to the X-ray spectrum of GRO J1655$-$40 (e.g. Miller et
  al.\ 2008).  However, the relative efficacy different models can
  still be easily gauged through the comparison of $\chi^{2}$ fit
  statistics.}
\vspace{-1.0\baselineskip}
\end{tiny}
\end{table}
\medskip

\clearpage

\begin{table}[t]
\caption{Strong Absorption Lines in MAXI J1305$-$704}
\begin{tiny}
\begin{center}
\begin{tabular}{ll}
\tableline
\tableline
Ion and Transition & Lab Wavelength \\
   ~               &      (\AA)     \\
\tableline
O VIII 1s--3p & 15.987 \\
O VIII 1s--2p & 18.9689 \\
Ne IX  1s$^{2}$--1s4p & 11.0003 \\
Ne IX 1s$^{2}$--1s3p & 11.5466 \\
Ne IX 1s$^{2}$--1s2p & 13.447 \\
Ne X 1s--6p & 9.3616 \\
Ne X 1s--5p & 9.4807 \\
Ne X 1s--4p & 9.7082 \\
Ne X 1s--3p & 10.2389 \\
Ne X 1s--2p & 12.1330 \\
Na XI 1s--2p & 10.0250 \\
Mg XII 1s--2p & 8.4210 \\
Mg XI 1s$^{2}$--1s2p & 9.1688 \\

\tableline
Fe XIX 2s$^{2}$2p$^{4}$--2s$^{2}$2p$^{3}$3d &   13.520 \\
Fe XX 2s$^{2}$2p$^{4}$--2p$^{2}$($^{3}$P)4d  &  9.991 \\
                  ~                        &  10.12 \\
Fe XX 2s$^{2}$2p$^{4}$--2p$^{2}$($^{3}$P)3d  &  12.82 \\
Fe XXI 2s$^{2}$2p$^{2}$--2s$^{2}$2p3d & 12.259 \\
                 ~                   & 12.325 \\
Fe XXII 2s2p$^{2}$--2s2p4d & 9.163 \\
            ~             & 9.183 \\
Fe XXII 2s$^{2}$2p--2s$^{2}$3d & 11.770 \\
            ~                 & 11.920 \\
Fe XXIII 2s$^{2}$--2s4p & 8.3029 \\
Fe XXIII 2s$^{2}$--2p3d & 10.175 \\
Fe XXIII 2s$^{2}$--2p3s & 10.560 \\
Fe XXIII 1s$^{2}$2s$^{2}$--1s$^{2}$2s3p & 10.980 \\
Fe XXIII 1s$^{2}$2s2p--1s$^{2}$2s3p    & 11.018 \\
Fe XXIV 1s$^{2}$2s--1s$^{2}$3p & 10.619 \\
            ~                 & 10.663 \\
\tableline
\tableline
\end{tabular}
\vspace*{\baselineskip}~\\ \end{center} 
\tablecomments{The table above lists lines that likely contribute
  strongly to the absorption spectrum observed in MAXI J1305$-$704 in
  the 8--19.5~\AA~ band, based on modeling with XSTAR photoionization
  calculations.  The quoted wavelengths are taken from Verner, Verner,
  \& Ferland (1996) and the NIST Atomic Spectrsocopy Database.  The
  transitions associated with the lines at 10.175~\AA~ and 10.560~\AA~
  should be regarded as tentative.  The density of lines in the MEG
  band considered in this analysis, the slightly degraded resolution
  of the MEG spectrum, and the intrinsic width of the lines serve to
  create some line blends.  While a self-consistent photionization
  model can handle this situation, line-by-line fitting with simple
  Gaussians cannot.  In addtion, some lines are saturated, meaning
  that velocity widths would effectively be upper-limits and line
  fluxes would be lower-limits.  Thus, individual line widths and
  fluxes are not quoted.  Relatively simple one- and two--zone
  photoionization models are described in Table 1.}
\vspace{-1.0\baselineskip}
\end{tiny}
\end{table}
\medskip

\clearpage

\centerline{~\psfig{file=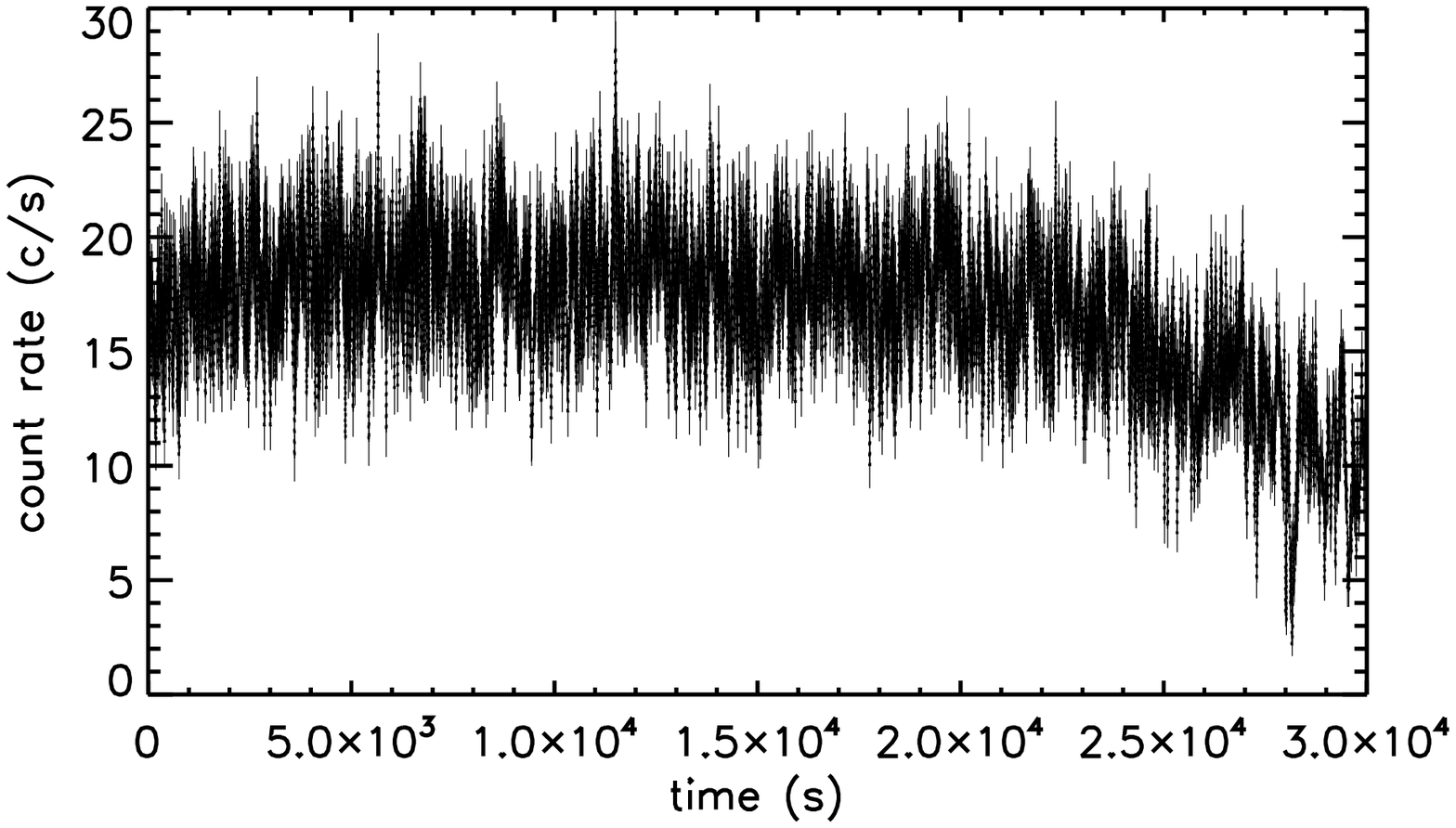,width=3.5in}~}
\figcaption[h]{\footnotesize The figure above shows the light curve of
  MAXI J1305$-$704, derived using 0.3--10.0~keV photons in the
  first-order MEG and HEG.  The time bins are 10~s (longer than the
  nominal 3.2~s ACIS frame time) in order to improve visual clarity.
  The light curve shows variability typical of accreting low-mass
  X-ray binaries, with slightly lower flux levels at the start and end
  of the observation.}
\medskip

\clearpage

\centerline{~\psfig{file=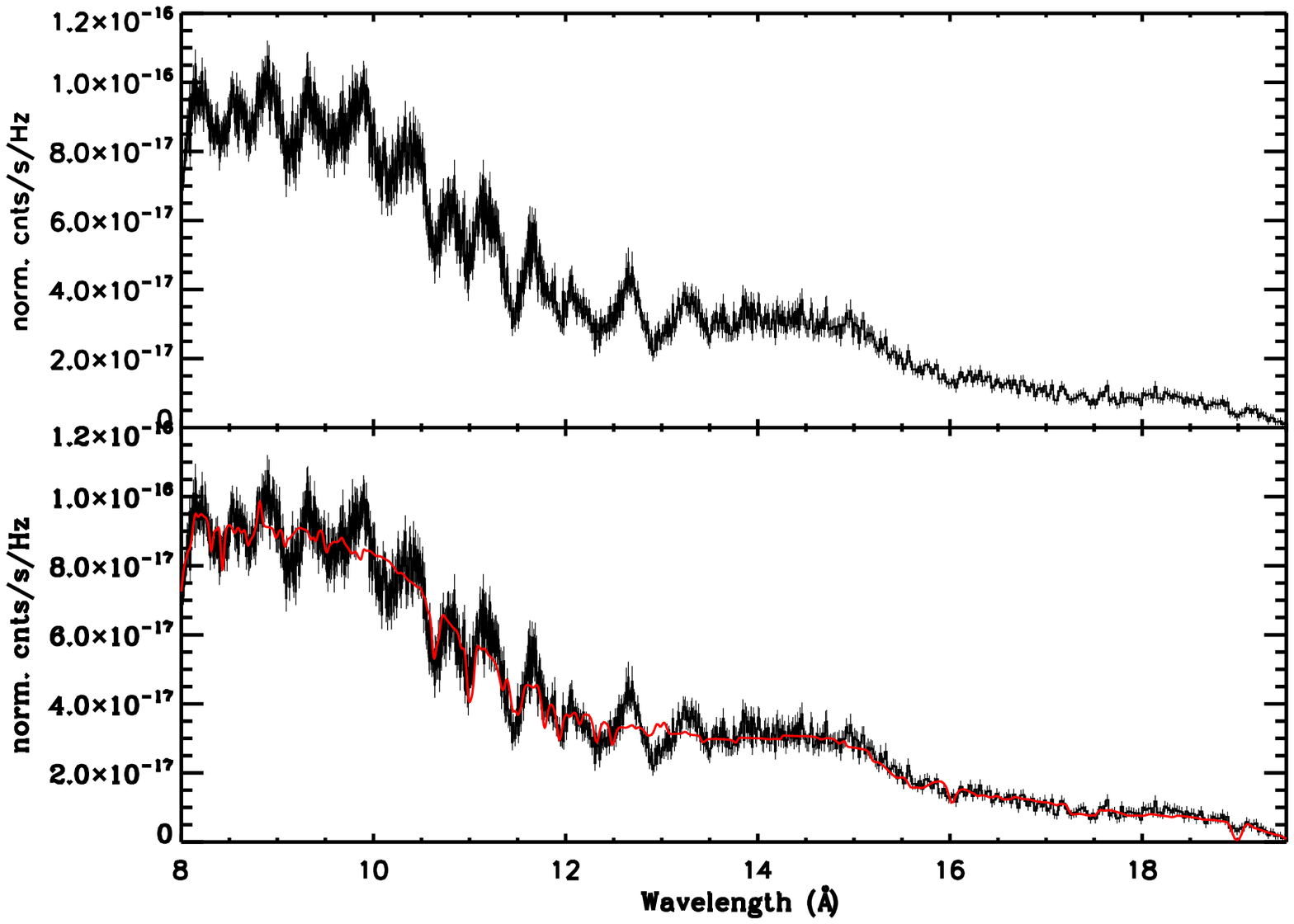,width=6.5in}~}
\figcaption[h]{\footnotesize The figure above shows the 8--19.5~\AA~
  combined first-order MEG spectrum of MAXI J1305$-$704.  The top
  panel shows the raw spectrum.  The bottom panel shows the best-fit
  two-zone absorption model, generated using a very large XSTAR grid
  (Model 16 in Table 1 is shown).  The plots are not ``fluxed'' to
  avoid bias, but the spectra have been binned for visual clarity.
  Absorption due to H-like O VIII is visible at 19~\AA~ (Ly-$\alpha$)
  and 16~\AA~ (Ly-$\beta$); most of the other features are Fe L-shell
  transitions.}
\medskip

\clearpage

\centerline{~\psfig{file=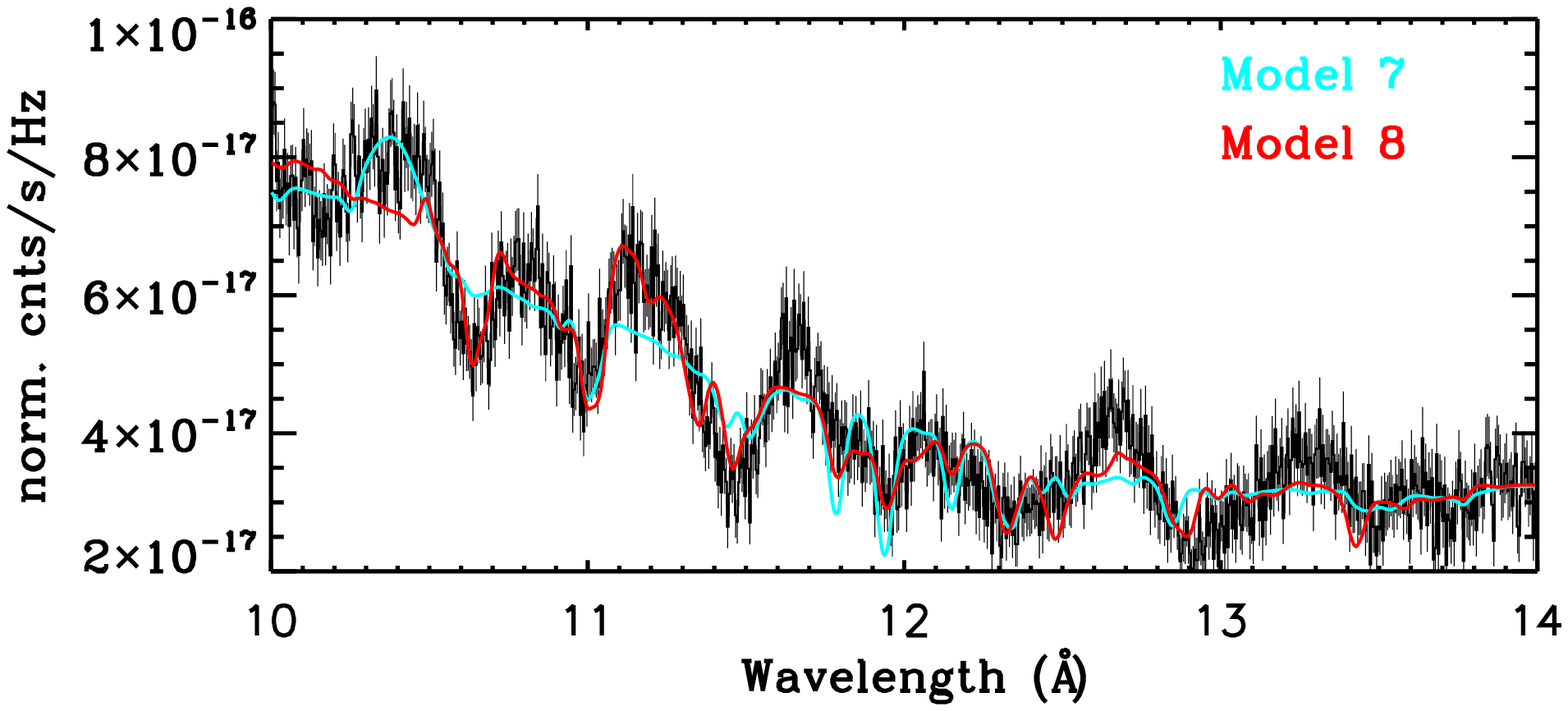,width=6.5in}~}
\figcaption[h]{\footnotesize The figure above shows the combined
  first-order MEG spectrum of MAXI J1305$-$704 in the 10--14~\AA~
  region, roughly centered on the density-sensitive Fe XXII lines
  (laboratory wavelengths: 11.77~\AA, 11.92~\AA).  Model 7 is a
  single-zone absorption model wherein the density is fixed at $n =
  1.0\times 10^{16}~ {\rm cm}^{-3}$ (see Table 1); it is plotted in cyan.  Model 8
  is a two-zone absorption model with variable (high) densities that
  yields a significantly improved fit; it is shown in red.  Note that
  the Fe XXII lines are fit much better with Model 8, as well as the
  lines at approximately 10.6~\AA, 11.5~\AA, and the complex near to
  12.9~\AA.  The plot has not been ``fluxed'' to
  avoid bias, but the spectra have been binned for visual clarity.}
\medskip

\centerline{~\psfig{file=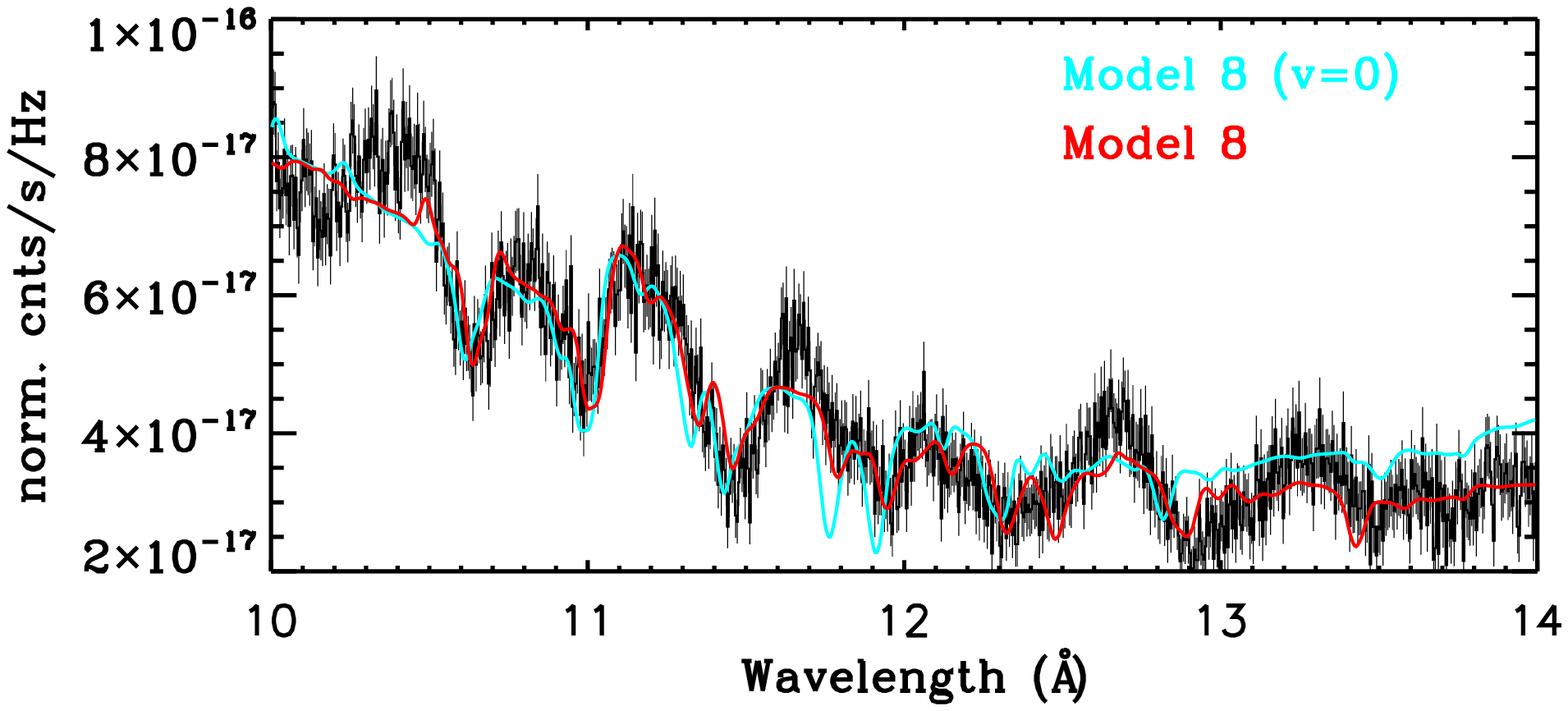,width=6.5in}~}
\figcaption[h]{\footnotesize The figure above shows the combined
  first-order MEG spectrum of MAXI J1305$-$704 in the 10--14~\AA~
  region, roughly centered on the density-sensitive Fe XXII lines
  (laboratory wavelengths: 11.77~\AA, 11.92~\AA).  Model 8 is shown in
  red as presented in Table 1.  It is also plotted in cyan after
  setting the velocity of both absorption components to zero.
  Clearly, the velocity shifts are important in obtaining robust fits
  to the spectrum.  The plot has not been ``fluxed'' to avoid bias,
  but the spectra have been binned for visual clarity.}
\medskip

\clearpage

\centerline{~\psfig{file=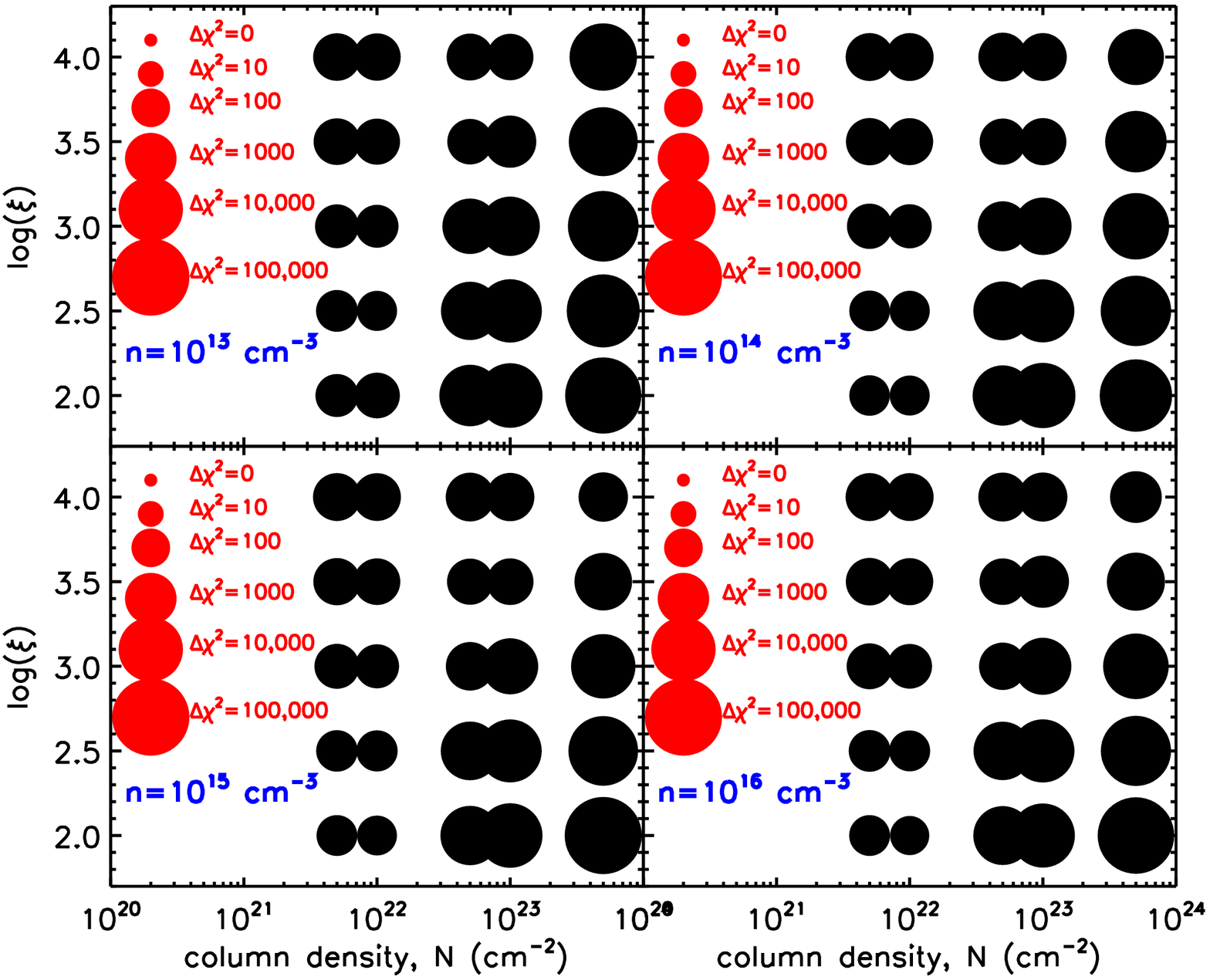,width=6.5in}~}
\figcaption[h]{\footnotesize The figure above shows the result of
  single--zone fits to the 10--14~\AA~ region using XSTAR
  photoionization models.  Each panel shows the results of fits with
  25 different combinations of ionization paramter $\xi$ and column
  density $N_{H}$, for a specific value of the gas density.  The size
  of the plotting symbol depicts a logarithmic scale in
  $\Delta(\chi^{2})$ versus the best--fit single--zone absorber (Model
  5 in Table 1; $n = 6\pm1 \times 10^{17}~ {\rm cm}^{-3}$).  This
  exploration of a broad three--dimensional parameter space shows that
  any fit with a low density gives $\Delta(\chi^{2}) > \sim100$.
  Indeed, in many cases, $\Delta(\chi^{2}) > 1000$.  This exercise
  demonstrates that a very high density is required to fit the
  spectrum of MAXI 1305$-$704.  }
\medskip

\clearpage

\centerline{~\psfig{file=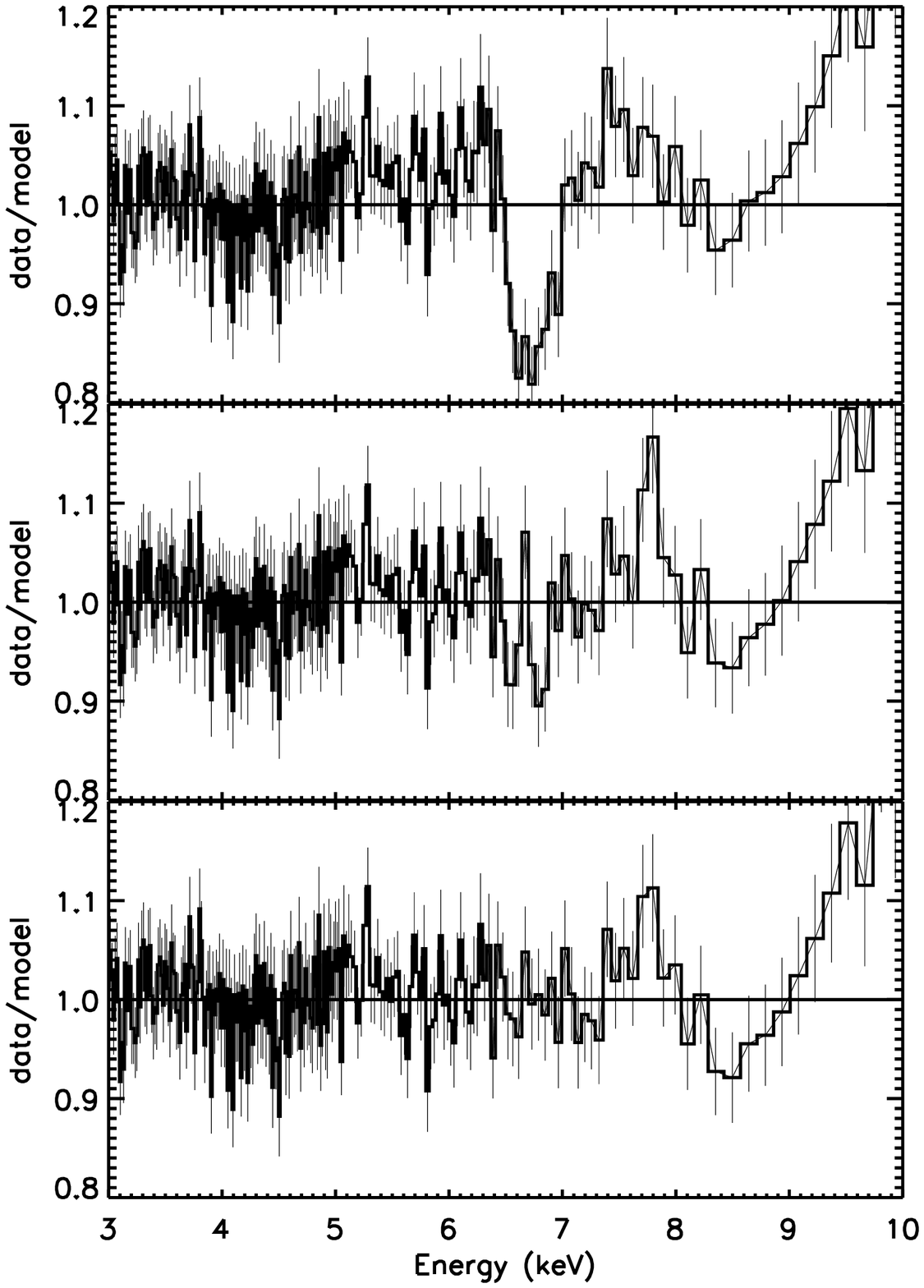,width=3.0in}~}
\figcaption[h]{\footnotesize The plot above shows the data/model
  ratios resulting from three fits to the zeroth-order ACIS spectrum
  of MAXI J1305$-$704.  The top panel shows a simple fit without
  absorption.  The middle panel shows a single-zone absorber with all
  parameters free, apart from $n = 1.0\times 10^{17}~ {\rm cm}^{-3}$
  (Model 14 in Table 1).  Note that residuals consistent with slightly
  red-shifted Fe XXV and XXVI remain.  The bottom panel shows a fit
  with two absorption zones; the second is significantly red-shifted
  (see Model 16 in Table 1).  Different tests find that the inclusion
  of a second, red-shifted absorber is significant at the 4--6$\sigma$
  level of confidence.  The data were binned for visual clarity.  The
  flux excess above 9~keV is due to pile-up.  In each case, the
  continuum model is fiducial.}
\medskip

\clearpage

\centerline{~\psfig{file=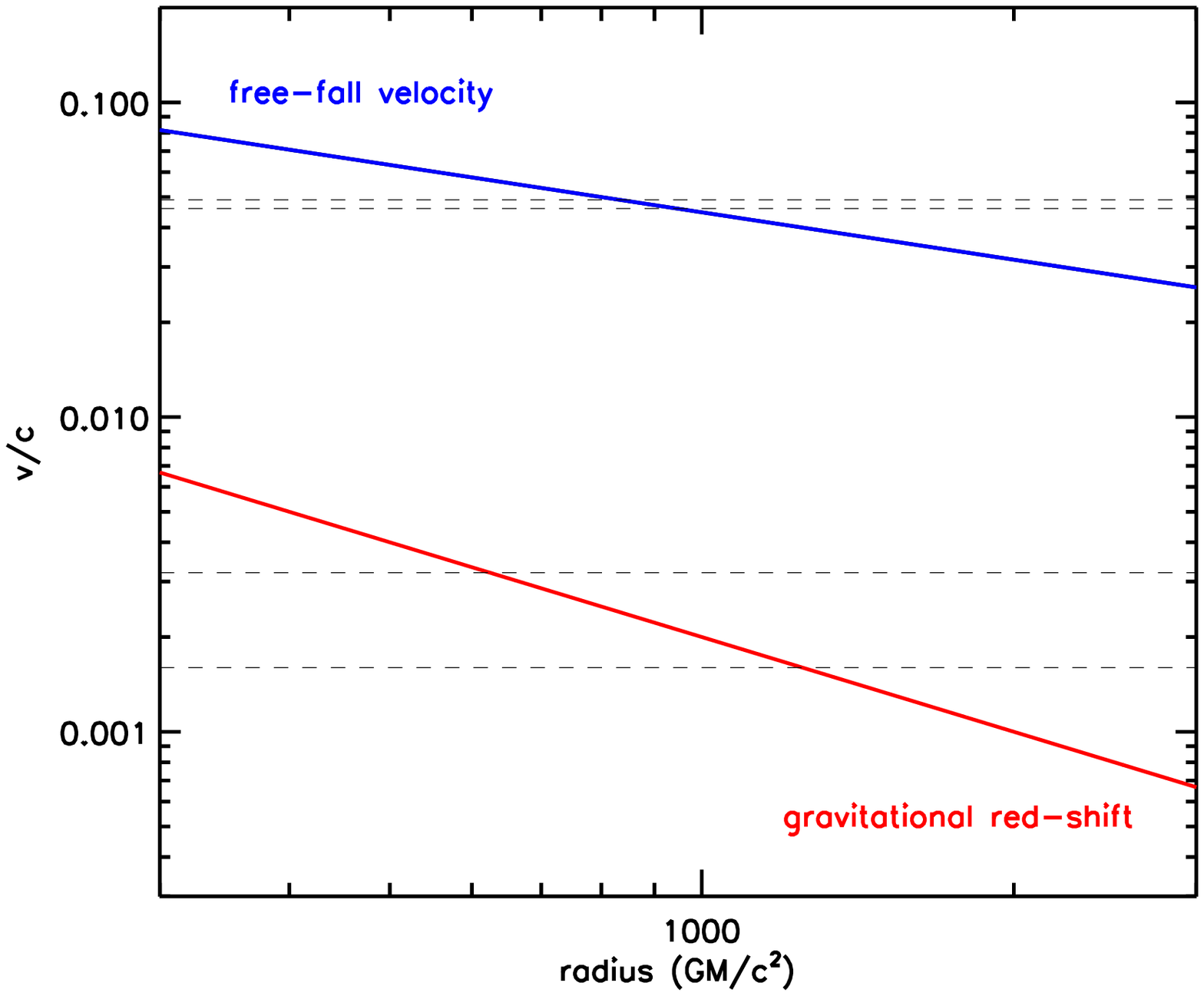,width=4.0in}~}
\figcaption[h]{\footnotesize The plot above shows the dependence of
  free--fall velocity (in blue) and gravitational red-shift velocity
  (in red), as a function of radius.  The dashed horizontal lines
  indicate the range of velocity shifts detected in fits to the MEG
  spectrum of MAXI J1305$-$704 via fits with both single--zone and
  two--zone XSTAR photoionization models with variable gas density
  parameters.  Those same photoionization models require that the
  absorption occur at a distance of $r = 3.9\pm 0.7 \times 10^{3}~
  {\rm km}$ from the compact object.  This radius corresponds to
  $1300~GM/c^{2}$ for a neutron star with $M = 2.0~M_{\odot}$; this
  only just intersects the range for which the two velocities can be
  observed at a common radius.  For black holes with masses of $M \leq
  5 M_{\odot}$, $r = 3.9\pm 0.7 \times 10^{3}$~km corresponds to radii
  for which the two velocities can be observed at a common distance
  from the central engine.  Other uncertainties, for instance in the
  luminosity, would allow for greater overlap in the radii implied by
  each velocity shift.
}
\medskip

\end{document}